\documentclass[sigconf]{acmart}
\AtBeginDocument{%
  \providecommand\BibTeX{{%
    \normalfont B\kern-0.5em{\scshape i\kern-0.25em b}\kern-0.8em\TeX}}}
\usepackage{xspace}
\usepackage{amsthm}
\usepackage{amsmath}
\usepackage{enumitem}
\usepackage{booktabs}
\usepackage{multirow}
\usepackage{graphicx}
\usepackage[normalem]{ulem}
\useunder{\uline}{\ul}{}
\usepackage{float}
\usepackage{caption}
\usepackage{subcaption}
\usepackage{balance}
\usepackage{threeparttable}
\usepackage{multicol}
\usepackage{multirow}
\usepackage[export]{adjustbox}
\usepackage{bm}
\newcommand{\modelname}{\textsf{ContrastVAE}\xspace}

\usepackage[T1]{fontenc}
\usepackage{aecompl}

\copyrightyear{2022}
\acmYear{2022}
\setcopyright{acmcopyright}
\acmConference[CIKM '22] {Proceedings of the 31st ACM International Conference on Information and Knowledge Management}{October 17--21, 2022}{Atlanta, GA, USA.}
\acmBooktitle{Proceedings of the 31st ACM International Conference on Information and Knowledge Management (CIKM '22), October 17--21, 2022, Atlanta, GA, USA}
\acmPrice{15.00}
\acmISBN{978-1-4503-9236-5/22/10}
\acmDOI{10.1145/3511808.3557268}

\begin{document}

\title[ContrastVAE: Contrastive Variational AutoEncoder for Sequential Recommendation]{ContrastVAE: Contrastive Variational AutoEncoder for Sequential Recommendation}

\author{Yu Wang}
\affiliation{%
  \institution{University of Illinois at Chicago}
 \city{Chicago}
  \country{United States}
}  
\email{ywang617@uic.edu}
\author{Hengrui Zhang}
\affiliation{%
  \institution{University of Illinois at Chicago}
 \city{Chicago}
  \country{United States}
}
\email{hzhan55@uic.edu}
\author{Zhiwei Liu}
\affiliation{%
  \institution{Salesforce}
 \city{San Francisco}
  \country{United States}
}
\email{zhiweiliu@salesforce.com}

\author{Liangwei Yang}
\affiliation{%
  \institution{University of Illinois at Chicago}
 \city{Chicago}
  \country{United States}
}
\email{lyang84@uic.edu}
\author{Philip S. Yu}
\affiliation{%
  \institution{University of Illinois at Chicago}
 \city{Chicago}
  \country{United States}
}
\email{psyu@uic.edu}

\renewcommand{\shortauthors}{Yu Wang et al.}

\begin{abstract}

Aiming at exploiting the rich information in user behavior sequences, sequential recommendation has been widely adopted in real-world recommender systems. However, current methods suffer from the following issues: 1) \textbf{sparsity} of user-item interactions, 2) \textbf{uncertainty} of sequential records, 3) \textbf{long-tail items}. In this paper, we propose to incorporate contrastive learning into the framework of Variational AutoEncoders to address these challenges simultaneously. Firstly, we introduce ContrastELBO, a novel training objective that extends the conventional single-view ELBO to two-view cases and theoretically builds a connection between VAE and contrastive learning from a two-view perspective. Then we propose Contrastive Variational AutoEncoder (ContrastVAE in short), a two-branched VAE model with contrastive regularization as an embodiment of ContrastELBO for sequential recommendation. We further introduce two simple yet effective augmentation strategies named \textit{model augmentation} and \textit{variational augmentation} to create a second view of a sequence and thus make contrastive learning possible. Experiments on four benchmark datasets demonstrate the effectiveness of \modelname and the proposed augmentation methods. Codes are available at \url{https://github.com/YuWang-1024/ContrastVAE}

\end{abstract}



\maketitle

\section{Introduction}
Sequential Recommendation (SR) has attracted increasing attention due to its ability to model the temporal dependencies in users' clicking histories, which can help better understand user behaviors and intentions.  
Recent research justifies the promising ability of self-attention models~\cite{SASRec, Bert4Rec, attn3} in characterizing the temporal dependencies on real-world sequential recommendation tasks. 
These methods encode a sequence as an embedding via an attention-based weighted sum of items' hidden representations.
To name a few,
SASRec~\cite{SASRec} is a pioneering work adopting the self-attention mechanism to learn transition patterns in item sequences, and Bert4Rec~\cite{Bert4Rec} extends it as a bi-directional encoder to predict the next item.

Despite their great representation power, both the \textit{uncertainty} problem and the \textit{sparsity} issue
impair their performance.

Firstly, the uncertainty problem is due to the rigorous assumption of sequential dependencies, which may be destroyed by unobserved factors in real-world scenarios. For example,  for music recommendations,
the genre of music that a user listens may vary according to different circumstances.
Nevertheless, those factors are unknown and cannot be fully revealed in sequential patterns.

Secondly, the sparsity issue is a long-existing and not yet a well-solved problem in recommender systems~\citep{liu2021augmenting, wang2021dskreg, yang2022large}. 
Supposing that a user only interacts with a few items, current methods are unable to learn high-quality representations of the sequences, thus failing to characterize sequential dependencies.
Moreover, the sparsity issue increases the deficiency of uncertainty in sequential recommendation. 
More concretely, if a user has fewer historical interactions, those uncertain factors are of higher dominance over sequential patterns. 
However, these two issues  are seldom studied simultaneously. 

Therefore, we investigate the potential of adopting Variational AutoEncoder (VAE) into sequential recommendation. The reasons are threefold. First of all, VAE can estimate the uncertainty of the input data. More specifically, it characterizes the distributions of those hidden representations via an encoder-decoder learning paradigm, which assumes that those representations follow a Gaussian distribution. 
Hence, the variances in Gaussian distribution can well characterize the uncertainty of the input data. Moreover, the decoder maximizes the expected likelihood of input data conditioned on such latent variables, which can thus reduce the deficiency from unexpected uncertainty. 
Secondly, the posterior distribution estimation in VAE decreases the vulnerability to the sparsity issue. Though a sequence contains few items, we can still characterize its distribution from learned prior knowledge and thus generate the next item. 
Last but not least, probabilistic modeling of those hidden representations also enhance the robustness of sparse data against uncertainty. 
Specifically, if we can ensure the estimated posterior of perturbed input still being in distribution, the decoder in VAE will tolerate such perturbations and yield correct next-item prediction.

However, conventional VAE suffers from posterior collapse issues~\cite{InfoVAE, MultVAE}. Concretely, if the decoder is sufficiently expressive, the estimated posterior distributions of latent factors tend to resemble the standard Gaussian distributions, i.e., these estimations are indistinguishable from each other as they follow the same distribution~\cite{MultVAE, C-DSVAE}. Furthermore, VAE might collapse to point estimation for rare classes that simply memorize the locations in latent space. The highly skewed distribution of user behaviors will exaggerate these problems. Specifically, the sequential input data consists of \textit{long-tail items}~\cite{CVAE}, which refer to the infrequent items that rarely appear in the users' historical records. Such items account for a large portion of all items. These limitations prevent VAE from achieving satisfactory performance for SR tasks.

Recent advances in adopting contrastive learning (CL) for alleviating representation degeneration problem~\cite{DuoRec} motivate us to resort to contrastive learning to mitigate the above issues. Concretely, contrastive learning encourages the uniform distribution of latent representations of different inputs~\cite{hypersphere}, thus enforcing them distinguishable in latent space. Besides, augmentations in CL encourage perturbed sequences to share similar representations, thus being robust to a large variance of the estimated posterior distribution. To incorporate contrastive learning into the framework of VAE, we first extend the conventional single-variable ELBO to the two-view case and propose ContrastELBO. We theoretically prove that optimizing ContrastELBO induces a mutual information maximization term, which could be effectively optimized with CL~\cite{CPC, mi_bound}. 

To instantiate ContrastELBO for SR, we propose \modelname, a two-branched VAE model that naturally incorporates CL. \modelname takes two augmented views of a sequence as input and follows the conventional encoder-sampling-decoder architecture to generate the next predicted item. The model is learned through optimizing an additional contrastive loss between the latent representations of two views in addition to the vanilla reconstruction losses and KL-divergence terms. To deal with the potential inconsistency problem led by uninformative data augmentations, we further propose two novel augmentation strategies: \textit{model augmentation} and \textit{variational augmentation}, which introduce perturbations in the latent space instead of the input space. We conduct comprehensive experiments on four benchmark datasets, which verify the effectiveness of the proposed model for sequential recommendation tasks, especially on recommending long-tail items. The contributions of this paper are summarized as follows:
\begin{itemize}
    \item We derive ContrastELBO, which is an extension of conventional single-view ELBO to two-view case and naturally incorporates contrastive learning into the framework of VAE.
    \item We propose ContrastVAE, a two-branched VAE framework guided by ContrastELBO for sequential recommendation.
    \item We introduce model augmentation and variational augmentation to avoid the semantic inconsistency problem led by conventional data augmentation.
    \item We conduct comprehensive experiments to evaluate our method. The results show that our model achieves state-of-the-art performance on four SR benchmarks. Extensive ablation studies and empirical analysis verify the effectiveness of the proposed components.
\end{itemize}

\section{Related Works}
\subsection{VAE for Sequential Recommendation}
Variational AutoEncoder (VAE) approximates the posterior distribution of latent variables given input data through variational inference and has been introduced to recommender systems recently to model the uncertainty of user preferences.  In the context of SR, SVAE~\cite{SVAE} and VSAN~\cite{VSAN} adopt VAE to learn the dynamic hidden representations of sequences by utilizing Recurrent Neural Network / Self-Attention as the Encoder and Decoder, respectively. Despite their abilities to model the uncertainty of sequential behaviors, empirical studies show that they perform worse than deterministic models on a lot of tasks~\cite{SASRec, Bert4Rec}, which is usually attributed to the low quality of approximate posterior distribution~\cite{ACVAE}. To alleviate such an issue, ACVAE~\cite{ACVAE} introduced adversarial variational Bayes combined with mutual information maximization between user embeddings and input sequences in order to obtain more salient and personalized representations of different users. However, DIM~\cite{DIM} has revealed that it is insufficient to optimize mutual information between latent representations and inputs merely, so we resort to latent-level contrastive learning through augmentations instead.

\subsection{Contrastive Sequential Recommendation}
Through maximizing the agreement of the representations of different augmentations of the input data, CL~\citep{CPC, DIM, SimCLR} has become a popular method for recommender systems in order to improve the accuracy and robustness of recommendation models~\cite{SGL_ED, DuoRec}. As a pioneering work, S3Rec~\cite{S3Rec} utilizes the correlations among attributes, items, and sub-sequences through mutual information maximization. CL4Rec~\cite{CL4Rec} introduces sequence masking, cropping, and reordering for sequential data augmentation and applies InfoNCE loss for item-level CL. To generate informative augmentations for user behavior sequences, CCL~\cite{CCL} employs a learnable context-aware generator for data augmentation, and then CL is performed on different augmented samples. 
To improve the robustness of data augmentations on item sequences, CoSeRec~\cite{CoSeRec} takes the correlations between different items into consideration. DuoRec~\cite{DuoRec} points out that the representations of items tend to become non-informative for many SR algorithms and highlights the limitations of random data-augmentation-based CL for SR. They propose to choose semantically similar sequences (with the same ground truth label) as their positive samples to address the inconsistency problem caused by random data augmentation. Different from data-augmentation-based models, our method works competitively with the simple dropout operation. We further introduce variational dropout~\cite{variational_dropout}, an adaptive dropout augmentation method to generate multiple views for a sequence, which naturally fits our VAE framework.

\section{Preliminary}
\subsection{Variational AutoEncoders and ELBO}\label{sec:singleview-elbo}
Given observed data $x$, latent variable models assume an underlying generative model $p(x,z) = p_{\theta}(x|z) p(z)$, where $z$ is the latent variable, $p(z)$ is the prior, and $p_{\theta}(x|z)$ is the likelihood parameterized by $\theta$. Then a natural objective is to maximize the log-likelihood of the data distribution:
\begin{equation}\label{eq:loglike}
   \log p(x) = \log \int_z p(z) p_{\theta}(x|z) \rm{d}z
\end{equation}
However, it is intractable to directly optimize Eq.~\ref{eq:loglike} as it requires integration over all possible $z$. To mitigate this issue, VAEs~\cite{VAE} adopt variational inference and use an approximate posterior distribution $q_{\phi}(z|x)$. Then $p_{\theta}(x|z)$ and $q_{\phi}(z|x)$ are jointly optimized through maximizing the Evidence Lower Bound (ELBO) to the log-likelihood:
\begin{equation}\label{eq:elbo}
\begin{split}
    \log p(x) \ge & \mathbb{E}_{q_{\phi}(z|x)}[\log p_{\theta}(x|z)] - D_{KL}[q_{\phi}(z|x)||p(z)]. \\
    = & \mathcal{L}_{ELBO}.
\end{split}
\end{equation}
In VAE, both $q_{\phi}(z|x)$ (encoder) and $p_{\theta}(x|z)$ (decoder) are parameterized by neural networks. In Eq.~\ref{eq:elbo}, the first term on the right is the expected conditional joint log-likelihood w.r.t the approximate posterior and is approximately computed through sampling with a reparameterization trick. The second term is the Kullback-Leibler divergence between the approximate posterior $q_{\phi}(z|x)$ and the prior $p(z)$, whose closed-form solution is easy to compute when assuming both are Gaussian distributions.

\subsection{Posterior collapse in VAE}
Despite the success of VAEs, a dominant issue named posterior collapse has been observed, which greatly reduces the capacity of the generative model~\cite{InfoVAE, collapse1, collapse2}. The posterior collapse is usually formulated as $D_{KL}[q_{\phi}(z|x) || p(z)] \rightarrow 0$ for every $x$, and usually occurs when the decoder model is too powerful. This indicates that the learned variational distribution is almost identical to the prior (i.e., standard Gaussian distribution), thus making the latent variables of different inputs indistinguishable in the latent space. ~\citet{InfoVAE} and ~\citet{collapse2} further point out that posterior collapse makes the mutual information between the input and its latent variable $\mathcal{I}(x, z) $ vanishingly small. 

To address this issue, recent methods try to reduce the impact of the KL-divergence term by reducing its weight~\cite{VSAN,MultVAE, RecVAE, SupervisedBetaVAE} or introducing an additional regularization term that explicitly maximizes the mutual information between the input and latent~\cite{InfoVAE, DuoRec, C-DSVAE}. 
However, this issue is much more serious in SR tasks as the user-item interactions are extremely sparse, and the user's dynamic preferences would be hard to model. 
Furthermore, we find that these methods are insufficient for better performance on the SR. 
As a remedy, we address the problem from the two-view CL perspective, where we maximize the mutual information between two views of each sequence in latent space $\mathcal{I}(z,z')$. 
In Section~\ref{sec:multiview-elbo} we show that through extending the vanilla single latent variable generative model to the two-view case, the VAE framework could naturally incorporate the mutual information maximization principle, which could be optimized through a CL loss~\cite{DIM, CPC,mi_bound}.

\section{Methodology}

\subsection{Problem Definition}
In SR tasks, there is a set of users $\mathcal{U} = \{ u_1, u_2 \dots, u_M\}$ and a set of items $\mathcal{V} = \{v_1, v_2, \dots, v_N\}$. For each user $u$, we are given its historical interaction records with the items, which are sorted according to their timestamps. Then in this case user $u$'s behaviour could be represented as an item sequence $\mathbf{X}^u = {x}^u_{1:T} = \{{x}_1^u, {x}_2^u,\dots, {x}_{T}^u\}$, where $x_t^u$ is the index of the item that user $u$ has interacted at time step $t$, and $T$ is the number of total time steps (i.e., the number of items user $u$ has interacted with). The target of SR is to predict the next item $x_{T+1}^u $ given the previous clicked items $x^u_{1:T}$. For each $x^u_t$, we further introduce a latent variable $z^u_t$ representing user $u$'s preference at time $t$. In Section~\ref{sec:multiview-elbo}, we  first derive the general formulation of ContrastELBO in Eq.~\ref{eq:elbo}, which extends the traditional single latent-variable ELBO to two-view case. We then discuss how to apply the ContrastELBO in SR scenarios in Section~\ref{sec:contrastvae}.

\subsection{ContrastELBO}\label{sec:multiview-elbo}
We start by introducing the following theorem about the latent variable model with two observed variables $x$ and $x'$, which represent two views of input:
\begin{theorem}
(ContrastELBO). Consider a generative model with two observed variables $(x, x')$ and two latent variables $(z, z')$ with structure $ x \leftarrow z - z' \rightarrow x'$, where $z$ is used to generate $x$, $z'$ is used to generate $x'$. $z$ and $z'$ are dependent as they are two views of the same input. This generative model indicates that joint distribution can be factorized by $p(z,z',x,x') = p(z,z')p(x|z)p(x'|z')$. Then we have the following lower bound of the log joint probability of the observed variables:

\begin{equation}\label{eqn:multiview-elbo}
\begin{split}
    \log p(x, x')  \ge & \quad \mathbb{E}_{q(z|x)}\log p(x|z)  - D_{KL}[q(z|x)||p(z)]  \\
    & + \mathbb{E}_{q(z'|x')}\log p(x'|z') - D_{KL}[q(z'|x')||p(z')]  \\
    & + \mathbb{E}_{q(z,z'|x,x')}\log\left[\frac{p(z,z')}{p(z)p(z')} \right]
\end{split}
\end{equation}
\end{theorem}

\begin{proof} According to the above generative model, we have $x$ and $x'$ that are conditionally independent give $z$ and $z'$, or formally $p(x,x'|z,z')$, then we can approximate the posterior with a variational distribution $q(z,z'|x,x')$ which could be factorized through:
\begin{equation}
    q(z,z'|x,x') = q(z|x)q(z'|x').
\end{equation}
Then we have
\begin{equation}\label{eqn:multiview-elbo-2}
\begin{split}
   \log p(x,x') & = \log\int p(x, x', z, z') dzdz'\\
   & = \log \mathbb{E}_{q(z,z'|x, x')} \left[\frac{p(x, x', z, z')}{q(z,z'|x, x')} \right]\\
   & \geq \mathbb{E}_{q(z,z'|x, x')}\log \left[\frac{p(x, x', z, z')}{q(z,z'|x, x')} \right]\\
   & = \mathbb{E}_{q(z,z'|x, x')}\log \left[\frac{p(x|z)p(x'|z')p(z,z')}{q(z|x)q(z'|x')} \right]\\
   & = \mathbb{E}_{q(z|x)} \log[p(x|z)]+ \mathbb{E}_{q(z'|x')}\log [p(x'|z')] \\
   & \quad + \mathbb{E}_{q(z,z'|x, x')}\log\left[\frac{p(z,z')}{q(z|x)q(z'|x')}\right].
\end{split}
\end{equation}
The last term in the last equation could be further expanded:
\begin{equation}\label{eqn:multiview-elbo-3}
\begin{split}
    & \mathbb{E}_{q(z,z'|x, x')}\log\left[\frac{p(z,z')}{q(z|x)q(z'|x')}\right] \\
    & = \mathbb{E}_{q(z,z'|x,x')} \log\left[\frac{p(z, z')p(z)p(z')}{q(z|x)q(z'|x')p(z)p(z')}\right]\\
    & = \mathbb{E}_{q(z,z'|x,x')} \log\left[\frac{p(z, z')}{p(z)p(z')}\right] \\
    & \quad + \mathbb{E}_{q(z,z'|x,x')} \log\left[\frac{p(z)p(z')}{q(z|x)q(z'|x')}\right]\\
    & = \mathbb{E}_{q(z,z'|x,x')} \log\left[\frac{p(z, z')}{p(z)p(z')}\right]\\
    & \quad -D_{KL}[q(z|x)||p(z)] - D_{KL}[q(z'|x')||p(z')] \\
\end{split}
\end{equation}
Plugging Eq.~\ref{eqn:multiview-elbo-3} into Eq.~\ref{eqn:multiview-elbo-2}, then we complete the proof.
\end{proof}
Note that the first four terms on the right of Eq.~\ref{eqn:multiview-elbo} are identical to that of the vanilla ELBO in Eq.~\ref{eq:elbo} and could be effectively optimized using traditional VAE models. The last term $\mathbb{E}_{q(z,z' | x, x')} log \frac{p(z, z')}{p(z)p(z')}$, however, is hard to compute. To make this term tractable, we follow the practice in~\citet{infonce_vae} that specifies $p(z, z') = q(z, z')$, $p(z) = q(z)$ and $p(z') = q(z')$ through choosing specific prior distributions, and then this term becomes $\mathbb{E}_{q(z,z'|x,x')} \frac{q(z, z')}{q(z), q(z)}$. If 
taking its expectation under the true data distribution $p_{true}(x, x')$, the last term becomes:
\begin{equation}\label{eq:elbo-mi}
    \mathbb{E}_{q(z,z')}\frac{q(z, z')}{q(z)q(z')} = {D}_{KL}[q(z,z')|| q(z)q(z')] = I(z,z').
\end{equation}
Eq.~\ref{eq:elbo-mi} indicates that we can maximize the mutual information between $z$ and $z'$ from $q(z,z')$. Note that $(z, z')$ are the encoder's output taking a pair of data $(x, x')$ as input, so the mutual information term can be efficiently estimated using its tractable lower bounds through CL ~\cite{mi_bound, MINE, CPC}.

In section~\ref{sec:contrastvae} we present \modelname, a direct instantiation of ContrastELBO, for sequential recommendation.

\subsection{ContrastVAE}\label{sec:contrastvae}
\begin{figure*}
    \centering
    \includegraphics[width=0.9\textwidth]{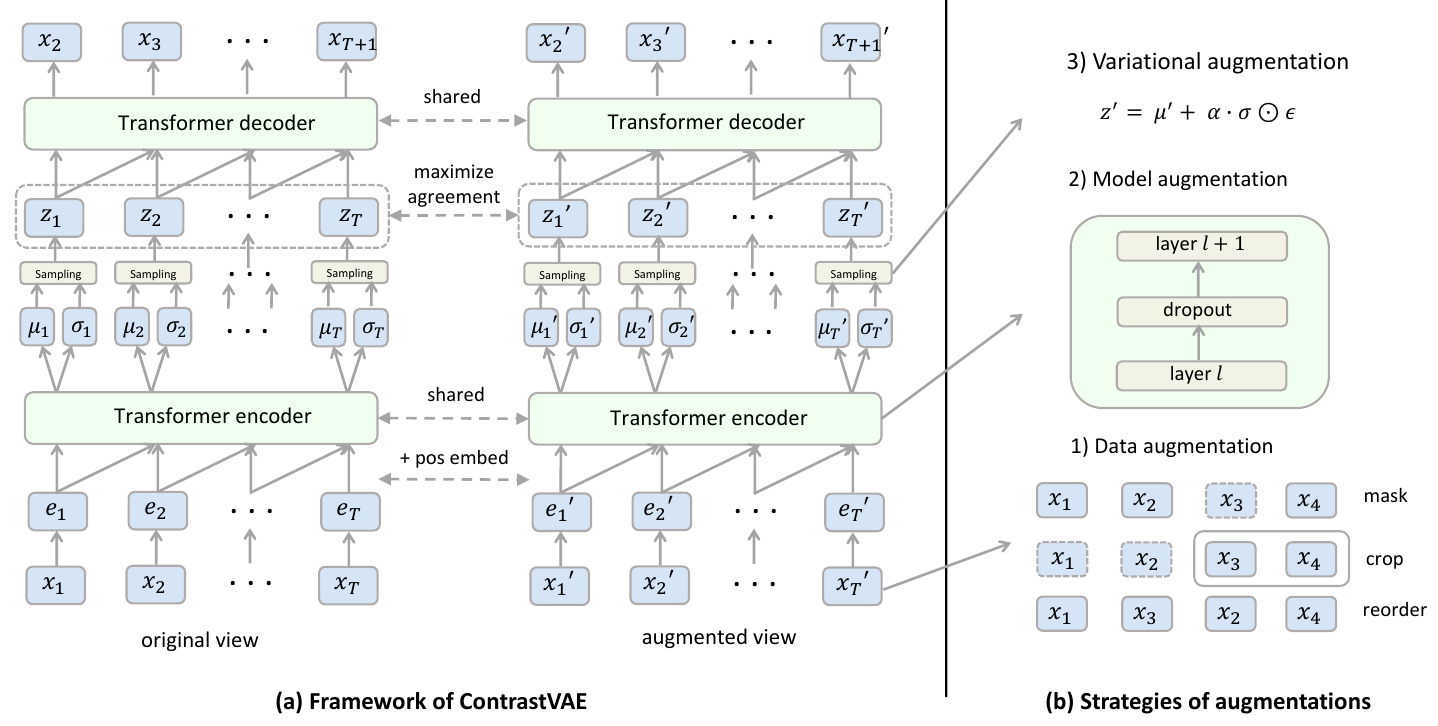}    
    \caption{A high-level illustration of \modelname. (a) The model framework: the model takes the original view $\bm{x}_{1:T}$ and an augmented view $\bm{x}_{1:T}'$ of the same sequence as input, and estimate the corresponding posterior distributions $q(z|x)$ and $q(z'|x')$ through a Transformer-based encoder. With sampled latent variables $\bm{z}_{1:T}$ and $\bm{z}_{1:T}'$, another Transformer-based decoder is used to perform next-item prediction tasks. The agreement of the latent variables of the two views is maximized through contrastive learning. (b) Strategies for generating the augmented view, where we consider three strategies that work at different steps of our model: 1) Data augmentation in input space, such as random masking, cropping, or reordering of a sequence; 2) Model augmentation, which performs random dropout in each intermediate layer of the Transformer encoder; 3) Variational augmentation, which introduces a learnable Gaussian dropout rate at sampling step.}
    \label{fig:framework}
\end{figure*}

The framework of ContrastVAE is presented in Fig.~\ref{fig:framework}. \modelname consists of three components: transformer encoder, augmentation strategies for generating the second view, and transformer decoder.
\subsubsection{Transformer encoder}\label{sec:encoder}
Inspired by the recent advances of transformer-based methods for SR~\cite{SASRec, Bert4Rec}, we adopt self-attention as the building block of our encoder. Note that our method is agnostic to the structure of the encoder and decoder, so other models such as RNN-based ones would work as well. The objective of the encoder is to infer the posterior distribution $q(z|x)$ (which is assumed to be a multivariate Gaussian distribution $\mathcal{N}({\bm{\mu}}, \bm{\sigma}^2 \mathbf{I})$), given the input sequence $x_{1:T}$:
\begin{equation}
\begin{split}
        q(z|x) &= q(\bm{z}_{1:T}|\bm{x}_{1:T})\sim \mathcal{N}(\bm{\mu}_{1:T}, \bm{\sigma}_{1:T}^2\mathbf{I})\\
      &=\prod_t q(\bm{z}_t|\bm{x}_{\leq t}) \sim \prod_t \mathcal{N}(\bm{\mu}_t, {\bm{\sigma}_t^2} \mathbf{I})\\
      & \bm{\mu}_t = \text{Enc}_{\mu}(\bm{x}_{\le t}) \\
      & \bm{\sigma}_t =  \text{Enc}_{\sigma}(\bm{x}_{\le t}),  \\
\end{split}
\end{equation}
where $\text{Enc}_{\mu}$ and $\text{Enc}_{\sigma}$ are two transformer-based encoders to estimate the corresponding mean $\bm{\mu}_{t}$ and covariance matrix $\bm{\sigma}_{t}^2\mathbf{I}$ of each input sequence. The detailed structure consists of the embedding layer and a stack of multiple self-attention layers. 
\paragraph{Embedding layer} To transform the discrete item indices into continuous vectors, we adopt a learnable item embedding table $\mathbf{M}\in \mathbb{R}^{N\times d}$, where $N$ is the total number of items and $d$ is the dimension of item embeddings. For any user's historical item sequence, we follow the common practice~\cite{SASRec} that first transforms it into a fixed-length of $T$, i.e., keeping the most recent $T$ items for sequences of length greater than $T$, and padding learnable tokens to the left of sequences whose lengths are smaller than $T$. Through the above operations we obtain the input sequence of a user, which can be represented as a matrix $\mathbf{E}=[\bm{e}_1, \bm{e}_2, \dots, \bm{e}_T]^{\top} \in \mathbb{R}^{T \times d}$, where $\bm{e}_t = \mathbf{M}_{{x}_t}$ is the embedding of item $x_t$ in the look-up table. Since the self-attention mechanism is not aware of the positions of tokens, we explicitly add additional position embeddings to the input embedding, which can be formalized as follows:
\begin{equation}
    \mathbf{\hat{E}} = \mathbf{E} + \mathbf{P} = 
    \begin{bmatrix}
    {\bm{e}_1} + \bm{p}_1 \\ {\bm{e}_2} + \bm{p}_2 \\...\\ {\bm{e}_T}+\bm{p}_T.
    \end{bmatrix},
\end{equation}
where $\bm{p}_n$ is the n-th position embedding and is a learnable vector. 
\paragraph{Self-attention layer}
The objective of SR is to recommend users of interest according to their historical interaction records. Previous methods like RNN~\cite{rnn2,rnn3} and Markov Chain~\cite{mc2, Fossil, FPMC} have issues of forgetting or short-range attention. The self-attention mechanism learns item representations attending to all items with adaptive weights, thus addressing the above issues. The input of the self-attention layer is $\mathbf{\hat{E}}$, and the operation can be formulated as follows:
\begin{equation}
    \text{SA}(\mathbf{\hat{E}}) = softmax\left(\frac{(\mathbf{\hat{E}}\mathbf{W}^Q)(\mathbf{\hat{E}}\mathbf{W}^K)^{\top}}{\sqrt{d}}\right) (\mathbf{\hat{E}}\mathbf{W}^V),
\end{equation} 
where $\mathbf{W}^Q$, $\mathbf{W}^K$ and $\mathbf{W}^V$ are the learnable weight matrices that map the sequence embedding to a different space. The self-attention operation computes item hidden representation as a weighted sum of all other item embeddings within this sequence, and the softmax function learns corresponding weights. $d$ is the dimension of latent item embeddings, and $\sqrt{d}$ is the normalization factor that avoids large values in the softmax function. To avoid information leakage and shortcut-learning, i.e., the model is aware of the items to be clicked in the future of time-step $t$, we mask all items after time step $t$ when computing the hidden representation of the sequence at $t$.

\paragraph{Stacking self-attention layers}
Since self-attention is the linear operation, we apply multi-layer perceptions over the output of self-attention layers with $relu$ as a non-linear activation function. Following~\citet{SASRec}, we also apply batch-normalization~\cite{batch_norm}, residual connection, and dropout as the model goes deeper. Finally, we stack multiple self-attention layers to form our encoder model $q(\bm{z}|\bm{x})$. This encoder takes the sequence embedding matrix as input and outputs the mean $\bm{\mu}_{1:T}$ and standard deviation $\bm{\sigma}_{1:T}$ of the amortized posterior Gaussian distribution. 

Then we can use the reparameterization trick to sample the latent variable $z$ accordingly:
\begin{equation}\label{eq:reparameterization}
    \bm{z}_{1:T} = \bm{\mu}_{1:T} + \bm{\sigma}_{1:T} \odot \bm{\epsilon},
\end{equation}
where $\epsilon \sim \mathcal{N} (0, \mathbf{I})$ is Gaussian noise.
 
\subsubsection{Transformer decoder}\label{sec:decoder}
The objective of the decoder is to learn the underlying generative process of data given the estimated distribution of latent factors and predict the next item of interest, which is reached by maximizing the reconstruction likelihood term in Eq.~\ref{eqn:multiview-elbo}:
\begin{equation}
\begin{split}
    \mathbb{E}_{q(\bm{z}|\bm{x})} \log[p(\bm{x}|\bm{z})] &= \mathbb{E}_{q(\bm{z}|\bm{x})} \log[p(\bm{x}_{1:T}|\bm{z}_{1:T})] \\
    & = \sum_{t} \mathbb{E}_{q(\bm{z}|x)}[\log p(\bm{x}_t|\bm{z}_{<t})].
\end{split}
\end{equation}
The decoder has almost the same structure as the encoder in Sec.~\ref{sec:encoder}, except that it does not require an embedding layer.

\subsubsection{Augmentation strategies}\label{sec-augmentation}
To facilitate \modelname with two-view input, we can use some augmentation strategies to create a second view for each input sequence. Although conventional data augmentation methods for sequential data, such as random cropping, shuffling, and masking~\cite{CL4Rec} can be naturally incorporated into our model, ~\citet{DuoRec} argues that they might lead to inconsistency problems between two augmented views, especially when the sequences are very short. To mitigate this issue, we resort to model-level augmentation methods that directly introduce randomness in the latent space.

Recent studies~\cite{simcse, DuoRec} show that the simple dropout operation is powerful enough to generate informative views for CL. And model augmentation~\cite{liu2022improving} is also studied to improve the sequential recommendation performance. These motivate us to merely use dropout operation at each intermediate layer of the encoder, which could be formulated as:
\begin{equation}\label{eq:dropout}
    \bm{h}^{(l+1)}_{1:T} = \bm{h}^{(l)}_{1:T} + \text{SA}(\bm{h}_{1:T}^{(l)} \odot \bm{\xi}) \text{ with }  \bm{\xi}_t \sim Bern(p),
\end{equation}
where $\bm{h}_{1:T}^{(l)}$ is the embedding at the $l$-th layer, $\bm{\xi}_t$ is $0/1$ masking sampled from Bernoulli distribution and $p$ is the dropout ratio. Note that we has added the residual connection term in Eq.~\ref{eq:dropout}.

Besides conventional Bernoulli dropout, Gaussian dropout~\cite{gaussian_dropout}, which multiplies output with random Gaussian noise, is an effective alternative with a faster convergence rate. Note that in the reparameterization trick in Eq.~\ref{eq:reparameterization}, we also multiply the standard deviation $\bm{\sigma}$ with random Gaussian noise to generate samples. This connection motivates us to adopt variational dropout~\cite{variational_dropout}, a special Gaussian dropout applied to the reparameterization step with a learnable dropout ratio:
\begin{equation}\label{eq:variational_dropout}
    \bm{z}_{1:T} = \bm{\mu}_{1:T} + \alpha \cdot \bm{\sigma}_{1:T} \odot \bm{\epsilon}
\end{equation}
where $\alpha$ is the learnable weight parameter to control the Gaussian dropout ratio.~\citet{variational_dropout} further interprets Eq.~\ref{eq:variational_dropout} from the Bayesian regularization perspective, where the KL divergence of the posterior distribution $q_{\alpha}(\bm{z}|\bm{x})$ (as a function of $\alpha$) and the prior $p(\bm{z})$ is minimized, if an additional regularization term of $\alpha$ is jointly optimized with the model:
\begin{equation}\label{eq:loss_vd}
    \mathcal{L}_{\alpha} = 0.5\log \alpha + 1.161 \alpha - 1.502\alpha^2 + 0.586\alpha^3. 
\end{equation}

In our implementation, dropout is equipped for both branches to create two views. The first branch always takes the original sequence (without augmentation) as input, and we term the input data and corresponding latent variable by $\bm{x}$ and $\bm{z}$. For the second branch, we apply additional strategies like data augmentations or variational augmentations to create a stronger augmented view. The input and corresponding latent variable is termed by $\bm{x'}$ and $\bm{z'}$. We provide an empirical study of different augmentation strategies in Sec.~\ref{sec:abl-aug}.

\subsection{Model Optimization}

With the model illustrated above, we can optimize the ContrastELBO presented in Sec.~\ref{sec:multiview-elbo}. In this section, we detailedly introduce our objective functions and how they correspond to different terms of the ContrarstELBO in Eq.~\ref{eqn:multiview-elbo}. Note that the optimizations of the reconstruction term and KL-divergence term for $(\bm{x}, \bm{z})$ and $(\bm{x}', \bm{z}')$ are identical and we only illustrate that for $(\bm{x},\bm{z})$.

To optimize the reconstruction term, we formalize it as a next-item prediction task where the log-likelihood could be factorized as follows:
\begin{equation}
    \log p(\bm{x}|\bm{z}) = \sum\limits_{t=1}^T \log p(\bm{x}_t | \bm{z}_{<t}).
\end{equation}
Denote the output of the encoder at position $t$ by $\mathbf{D}_t$, and the embedding of item $i$ by $\mathbf{M}_i$, then we define the probability that the next item is item $i$ given the hidden state before time $t$ $z_t$ as:
\begin{equation}
    p(\bm{x}_{t+1} = i | \bm{z}_{\le t}) \propto \mathbf{D}^{\top}_t \mathbf{M}_i.
\end{equation}
For each time step $t$, we denote the ground truth next item by $\mathbf{M}_t$. Then following ~\citet{SASRec}, we randomly sample another item as the negative example $\mathbf{M}_n$. Thus we can optimize the expected log-likelihood term by minimizing the cross-entropy loss:
\begin{equation}\label{eq:loss-ce}
        \mathcal{L}_{CE} = \sum_{t = 1}^{\top} \left[\log(\sigma (\mathbf{D}_t^T\mathbf{M}_t)) + 
    \log(\sigma(1-\mathbf{D}_t^{\top}\mathbf{M}_n))\right],
\end{equation}
where $\sigma$ is the sigmoid function.

For the KL-divergence term, we assume the prior $p(z) = \prod\limits_t p(z_t)$ and $p(z_t) = \mathcal{N}(\bm{0}, \bm{I})$ for all $t$. Combining with the encoder's design in Sec.~\ref{sec:encoder} that $q(z|x)$ is another Gaussian distribution $\mathcal{N}(\bm{\mu}, \bm{\sigma}^2\bm{I})$, the closed-form solution of the KL-divergence could be easily computed through:
\begin{equation}\label{eq:loss-kl}
\begin{split}
    \mathcal{L}_{KL} = & D_{KL}[q(\bm{z}|\bm{x}) || p(\bm{z})] \\
    = & \sum\limits_{t=1}^T D_{KL}[q(\bm{z}_t|\bm{x}_t) || p(\bm{z}_t) ] \\
    = & \sum\limits_{t=1}^T \sum\limits_{d=1}^D (\sigma_{t,d}^2 + \mu_{t,d}^2 - 1 - \log \sigma_{t,d}^2 ) \\
\end{split}
\end{equation}

To maximize the mutual information term between $\bm{z}$ and $\bm{z}'$ under $q(\bm{z},\bm{z}')$, we adopt the InfoNCE loss function~\cite{CPC, SimCLR}, which is a multi-sample unnormalized lower bound of mutual information with low variance~\cite{mi_bound}. Denote $u$'s hidden representation by $\mathbf{z}_u$, and its positive sample by $\mathbf{z}_u'$, then the loss could be formalized as:
\begin{equation}\label{eq:loss-mi}
    \mathcal{L}_{InfoNCE} = \frac{1}{M}\sum\limits_{u=1}^M  \log \frac{\exp(\mathbf{z}_{u}^{\top} \mathbf{z}'_u / \tau)}{\sum\limits_{v} \exp(\mathbf{z}^{\top}_u \mathbf{z}'_v / \tau) + \sum\limits_{v \neq u} \exp(\mathbf{z}^{\top}_u \mathbf{z}_v / \tau)},
\end{equation}
where $\tau$ is the temperature hyperparameter, $\mathbf{z}_u = \mathcal{R}(\bm{z}^u_{1:T})$ is a representation of the $i$-th sequence, which summarizes the latent representations of all tokens of the sequence. We use the simple average pooling to implement $\mathcal{R}(\cdot)$ in this paper.

Plugging Eqs.~\ref{eq:loss-ce},~\ref{eq:loss-kl},~\ref{eq:loss-mi} into the ContrastELBO in Eq.~\ref{eqn:multiview-elbo}, we get the final objective function (note that Eq.~\ref{eq:loss_vd} is required as well when using variational dropout):
\begin{equation}
\begin{split}
    \mathcal{L} = & \quad \mathcal{L}_{CE}  - \mathcal{L}_{KL} \quad \text{(ELBO of the 1st view)} \\
      + & \quad \mathcal{L}_{CE}' -  \mathcal{L}_{KL}' \quad \text{(ELBO of the 2nd view)} \\
      + &\quad \lambda \cdot \mathcal{L}_{InfoNCE} \quad \text{(InfoNCE)} \\
\end{split}
\end{equation}
where $\lambda$ is the weight of InfoNCE loss as Eq.~\ref{eq:loss-mi} is an unnormalized estimation of mutual information and does not reflect its magnitude.

\section{Experiment}
In this section, we evaluate the proposed \modelname empirically on real-world SR tasks, and we would like to answer the following research questions:
\begin{itemize}
    \item \textbf{RQ1}: How does \modelname perform compared with state-of-the-art SR models? 
    \item \textbf{RQ2}: Are the key components in \modelname, such as augmentations and contrastive learning, necessary and beneficial for satisfactory improvement?
    \item \textbf{RQ3}: How is the performance of \modelname on items with different frequencies and sequences with different lengths? Does \modelname improve the performance on long-tail items, and what are the reasons?
    \item \textbf{RQ4}: How is the robustness of \modelname w.r.t. noisy input sequences, and is \modelname sensitive to some key model hyperparameters?
\end{itemize}
\subsection{Setups}
\textbf{Dataset.} We conduct experiments on four datasets collected from Amazon review in various domains~\cite{amazon}:  \textit{Beauty}, \textit{Toys and Games (Toys)}, \textit{Tools and Home (Tools)} and \textit{Office Product (Office)}. We treat all the user-item rating records as implicit feedback and sort them according to the timestamps to form sequences. Following the common settings~\cite{SASRec}, we filter out users and items with less than five interaction records. 
For each user, we use the last clicked item for testing, the penultimate one for validation, and the remaining clicked items for training. We provide the statistics of the four datasets in Table~\ref{tab:data_stat}. Furthermore, we split the sequences into five groups according to the latest interacted item's frequency, and report the population of each group in Table~\ref{tab:data_freq_per_item}, where we can observe that most items lie in the smallest frequency group and exhibits a long-tail distribution.
\begin{table}[H]
\centering
\caption{Statistics of datasets, we report the number of users, number of items, number of interactions, number of interactions per item, and the averaged sequence length.}
\label{tab:data_stat}
\small
\begin{tabular}{@{}c|ccccc@{}}
\hline
Dataset& \#Users &\#Items &\#Interactions &\#Ints / item & Avg. seq. len.\\
\hline
Beauty & 22,363 & 12,101 & 198,502 & 16.40 & 8.3 \\
Toys & 19,412 & 11,924 & 167,597 & 14.06 & 8.6 \\
Tools & 16,638 & 10,217 & 134,476 & 13.16 & 8.1 \\
Office & 4,905 & 2,420 & 53,258 & 22.00 & 10.8 \\
\hline
\end{tabular}
\end{table}
\begin{table}[h]
    \centering
    \caption{Number of sequences end at items of different frequency groups.}
    \setlength{\tabcolsep}{2.6mm}{
    \begin{tabular}{c|ccccc}
         \hline
         Dataset & [$\leq$10] & [10, 20] & [20, 30] & [30, 40] & [$\geq$40]  \\
         \hline
         Beauty & 17,353 & 3,152 & 1,065 & 367 & 426\\
         Toys & 16,345 & 2,320 & 476 & 130 & 141\\
         Tools & 13,929 & 1,769 & 400 & 230 & 310\\
         Office & 3,150&  1,028 & 547 & 97 & 83\\
         \hline
    \end{tabular}
    }
    \label{tab:data_freq_per_item}
\end{table}

\begin{table*}[t]
    \caption{Overall Comparison}
    \label{tab:overall_comp}
    \small
    \centering
    \setlength{\tabcolsep}{1.8mm}{
    \begin{tabular}{l|l|cccccccccccc}
         \hline
         Dataset & Metric & SVAE & ACVAE & S3Rec & CL4Rec & LightGCN & BPRMF & Bert4Rec & SASRec & STOSA & DT4SR & \modelname & Improv.\\
         \hline
         \multirow{4}{*}{Beauty}&R@20& 0.0268 & 0.0951 & 0.0946 & 0.0398& 0.0759 & 0.0739 & 0.0890 & 0.0952 & 0.0975 & \underline{0.0982} & \textbf{0.1095} & 11.51\%\\
         & R@40 & 0.0417 & 0.1294 & 0.1348 & 0.0554& 0.1112 & 0.1089 & 0.1285 & 0.1389 & 0.1337 & \underline{0.1404} & \textbf{0.1541} & 9.76\%\\
         & N@20 &0.0102 & 0.0467& 0.0424 &  0.0168& 0.0306 & 0.0311 & 0.0395 & 0.0420 & \underline{0.0469} & 0.0446 & \textbf{0.0496} & 5.76\%\\
         & N@40 & 0.0132 & 0.0537 & 0.0505 & 0.0200& 0.0378 & 0.0383 & 0.0476 & 0.0509 & \underline{0.0542} & 0.0533 & \textbf{0.0587} & 8.30\%\\
         \hline
        \multirow{4}{*}{Office} & R@20 & 0.0988 & 0.1327 & 0.1335 & 0.0646 & 0.0532 & 0.0483 & 0.1350 & 0.1478 & \underline{0.1578} & 0.1429 & \textbf{0.1708} & 8.24\%\\
         & R@40 & 0.1647 & 0.2075 & 0.2112 & 0.1025 & 0.0797 & 0.0718 & 0.2230 & 0.2251 & \underline{0.2391} & 0.2186 & \textbf{0.2617} & 9.45\%\\
         & N@20 & 0.0389 & 0.0560 & 0.0571 & 0.0291 & 0.0243 & 0.0218 & 0.0551 & 0.0657 & \underline{0.0694} & 0.0643 & \textbf{0.0741} & 6.77\%\\
         & N@40 & 0.0523 & 0.0713 & 0.0729 & 0.0368 & 0.0297 & 0.0266 & 0.0729 & 0.0815 & \underline{0.0859} & 0.0797 & \textbf{0.0925} & 7.68\%\\
         \hline
        \multirow{4}{*}{Toy} & R@20 & 0.0178 & 0.0722 & 0.0973 & 0.0392& 0.0671 & 0.0692 & 0.0699 & 0.1112 & 0.1008 & \underline{0.1130} & \textbf{0.1164} & 3.01\%\\
         & R@40 & 0.0260 & 0.1030 & 0.1307 & 0.0596 & 0.0977 & 0.1007 & 0.0982 & \underline{0.1479} & 0.1357 & 0.1478 & \textbf{0.1610} & 8.86\%\\
         & N@20 & 0.0069 & 0.0359 & 0.0467 & 0.0182 & 0.0287 & 0.0304 & 0.0318 & \underline{0.0539} & 0.0496 & 0.0515 & \textbf{0.0547} & 1.48\%\\
         & N@40 & 0.0086 & 0.0421 & 0.0536 & 0.0224& 0.0349 & 0.0369 & 0.0376 & \underline{0.0614} & 0.0567 & 0.0560 & \textbf{0.0638} & 4.42\%\\
         \hline
        \multirow{4}{*}{Tool} & R@20 & 0.0340 & 0.0537 & 0.0632 & 0.0443 & 0.0537 & 0.0505 & 0.0508 & \underline{0.0640} & 0.0615 & 0.0601 & \textbf{0.0731} & 14.21\%\\
         & R@40 & 0.0521 & 0.0759 & 0.0849 & 0.0634 & 0.0751 & 0.0715 & 0.0777 & \underline{0.0879} & 0.0867 & 0.0861 & \textbf{0.1049} & 19.34\%\\
         & N@20 & 0.0149 & 0.0249 & 0.0286 & 0.0194 & 0.0238 & 0.0219 & 0.0213 & 0.0294 & \underline{0.0295} & 0.0289 & \textbf{0.0326} & 10.51\%\\
         & N@40 & 0.0186 & 0.0294 & 0.0330 & 0.0233 & 0.0282 & 0.0262 & 0.0268 & 0.0345 & \underline{0.0346} & 0.0342 & \textbf{0.0381} & 10.12\%\\
         \hline
    \end{tabular}
    }
\end{table*}
\textbf{Metrics.} 
We compute each user's relevance scores for all items and choose items with top-$N$ scores for the recommendation. Then we adopt two widely used top-$N$ metrics, Recall and NDCG, as our top-N ranking evaluation metrics. We report the experimental results when $N=20$ and $N=40$. 

\textbf{Baselines.}
We compare our methods with four types of representative SR models: 1) VAE-based methods, including SVAE~\cite{SVAE} and ACVAE~\cite{ACVAE}; 2) CL-based methods, including S3Rec~\cite{S3Rec} and CL4Rec~\cite{CL4Rec}; 3) attention-based methods, including Bert4Rec~\cite{Bert4Rec}, SASRec~\cite{SASRec}; \textit{4) probabilistic SR}, including STOSA~\cite{STOSA}, DT4SR~\cite{DTS4R}. We also compare our methods with collaborative filtering methods that ignore sequential information: LightGCN~\cite{LightGCN} and BPRMF~\cite{BPR}.

\textbf{Implementation Details.}
We use PyTorch to implement our model, and all experiments are conducted on an Nvidia V100 GPU with 16G memory. We use Adam~\cite{adam} to optimize our method. To avoid overfitting, we adopt an early stop strategy to stop experiments when there is no improvement in $100$ epochs. We set the learning rate as $0.001$, hidden dimension as $128$, model dropout probability as $0.3$, and the number of attention heads as $4$ for all models.

\subsection{Comparative Results}
We report the performance of \modelname and other comparative methods in Table~\ref{tab:overall_comp}. Our model \modelname consistently outperforms other
methods throughout the four datasets. Concretely, \modelname achieves $8.86\%$- $19.34\%$ improvements of Recall@40 and $4.42\%$ - $10.12\%$ improvements of NDCG@40 compared with the best baseline, which shows the effectiveness of our method for the SR. Furthermore, we find that on the Tool dataset, which has the smallest number of interactions per item and per user, our method outperforms the most powerful baselines by over $19\%$. We attribute this to the better capability of our model to learn from the noise and uncertainty of user behaviors, which is especially beneficial for short sequences (sparsity) and long-tail items.

\subsection{Ablation Studies}
In order to verify the effectiveness and necessities of the critical components and compare the performance of different designs of our method, we conduct several ablation studies for \modelname.

\subsubsection{Effect of CL and variational dropout}
First of all, to validate the importance of CL and the regularization loss in variational dropout, we consider two variants of \modelname: 1) removing the mutual information term (i.e., the InfoNCE loss in Eq.~\ref{eq:loss-mi}), 2) removing the regularization term of the variational dropout rate $\mathcal{L}_{\alpha}$ in Eq.~\ref{eq:loss_vd}. We present the results of Recall@20 and NDCG@20 on Toy and Tool datasets in Table.~\ref{tbl-abl-contrast}.
The results show that both the CL and the regularization of variational dropout are important to our model, and the model performance will degrade greatly if they are removed.
\begin{table}[H]
	\centering
	\caption{Performance of \modelname when removing the CL loss (w/o MI) and regularization of variational dropout (w/o $\mathcal{L}_{\alpha}$).}
	\begin{adjustbox}{width=\columnwidth,center}
	\small
	\label{tbl-abl-contrast}
	\begin{threeparttable}
        {
		\begin{tabular}{lcccccc}
			\toprule[0.8pt]
			\multirow{2}{*}{{Methods}}   & \multicolumn{2}{c}{Toy} & \multicolumn{2}{c}{Tool} \\
		    \cline{2-3} \cline{4-5}
            & R@20 & N@20  & R@20 & N@20  \\
	        \midrule
            w/o MI  & 0.112 (-3.60\%)  & 0.052 (-5.85\%) & 0.070 (-4.11\%)  & 0.031 (-5.83\%)  \\
            w/o $\mathcal{L}_{\alpha}$  & 0.104 (-10.9\%)  & 0.051 (-6.03\%) & 0.063 (-13.69\%) & 0.028 (-14.4\%) \\
            \midrule
            default & \textbf{0.116} & \textbf{0.055} & \textbf{0.073} & \textbf{0.033}\\
			\bottomrule[0.8pt]
		\end{tabular}}
	\end{threeparttable}
	\end{adjustbox}
\end{table}
\subsubsection{Comparison of different augmentation strategies}\label{sec:abl-aug} To evaluate the effectiveness of different augmentation methods, we compare the performance when equipping different augmentation strategies to \modelname, including data augmentation (DA), model augmentation (MA), and variational augmentation (VA): 1) For data augmentation, we follow~\citet{CL4Rec} and adopt random cropping, masking and reordering together to generate a perturbed version for each input sequence, and use it as the input of the second branch of our model; 2) For model augmentation, we simply apply the basic dropout to the encoder of the second branch; 3) For variational augmentation, we apply variational dropout at the sampling step with reparameterization as introduced in Sec.~\ref{sec-augmentation}. As shown in Table~\ref{tab_abl_augmentation} all three augmentation strategies improve the model's performance by a large margin, compared with the baseline method AVAE, which merely uses single-branch VAE without CL. We also have the following interesting observations: 1) compared with DA, which achieves the best performance on $2$ out of $16$ metrics, MA and VA are better on more metrics and datasets ($6$ out of $16$ and $8$ out of $16$ respectively), and this demonstrates the limitations of DA, which will introduce inconsistency between augmented views. 2) DA performs best on the Office dataset. This might be because the Office dataset has the largest average sequence length (see Table~\ref{tab:data_stat}) and thus is less sensitive to the perturbation of data augmentations. 3) MA method performs competitively compared with DA but is much simpler. 4) Our proposed VA achieves comparable or even better results, especially on Toy and Tool datasets with smaller average sequence length and testing item frequency. This shows that the proposed VA can effectively benefit the prediction of short sequences and long-tail items.
\begin{table}[H]
\vspace{-8pt}
    \caption{Comparison of the performance of different augmentation strategies on the four datasets: AVAE is short for the single-branch VAE model that uses an attentive encoder and decoder without CL. DA is short for data augmentation, MA is short for model augmentation, and VA is short for variational augmentation.}
    \small
    \centering
    \label{tab_abl_augmentation}
    \setlength{\tabcolsep}{3.0mm}
    {
    \begin{tabular}{l|l|cccccccccccc}
         \hline
         Dataset&Metric & AVAE &  DA & MA & VA \\
         \hline
         \multirow{4}{*}{Beauty} & R@20 & 0.0448  & 0.1059 &  \textbf{0.1095} & 0.1066\\
         & R@40 & 0.0709 & 0.1561  & 0.1541 & \textbf{0.1578} \\
         & N@20 & 0.0180 & 0.0459 & \textbf{0.0496} & 0.0464 \\
         & N@40 & 0.0233 & 0.0562 & \textbf{0.0587} & 0.0568 \\
         \hline
        \multirow{4}{*}{Office} & R@20 & 0.1093 & \textbf{0.1745} & 0.1708 & 0.1672 \\
         & R@40 & 0.1918 & \textbf{0.2658} & 0.2617 &  0.2599 \\
         & N@20 & 0.0419 & 0.0739 & \textbf{0.0741} & 0.0722 \\
         & N@40 & 0.0586 & {0.0924} & \textbf{0.0925} & 0.0911 \\
         \hline
        \multirow{4}{*}{Toy} & R@20 & 0.0423 & 0.1112 & 0.1130  & \textbf{0.1164} \\
         &R@40 & 0.0700 & 0.1554 & 0.1548  & \textbf{0.1610} \\
         &N@20 & 0.0171  & 0.0503 & 0.0566 & \textbf{0.0547} \\
         &N@40 & 0.0227 & 0.0593 & \textbf{0.0652}  & 0.0638 \\
         \hline
        \multirow{4}{*}{Tool} & R@20 & 0.0380 & 0.0671 & 0.0691 & \textbf{0.0731} \\
         &R@40 & 0.0603 & 0.1004  & 0.0986 & \textbf{0.1049}\\
         &N@20 & 0.0164 & 0.0295  & 0.0310 & \textbf{0.0326}\\
         &N@40 & 0.0209 & 0.0364  & 0.0370 & \textbf{0.0381}\\
         \hline
    \end{tabular}
    }
\end{table}

\subsection{Analysis of \modelname}\label{sec:analysis}
In this section, we study the strengths of the proposed \modelname by analyzing its property from several aspects.

\subsubsection{\modelname benefits the prediction of long-tail items and for short sequences}\label{sec:study1}
\begin{figure}[h]\label{fig:sub_performance}
\begin{subfigure}{0.48\columnwidth}
\includegraphics[width=\linewidth]{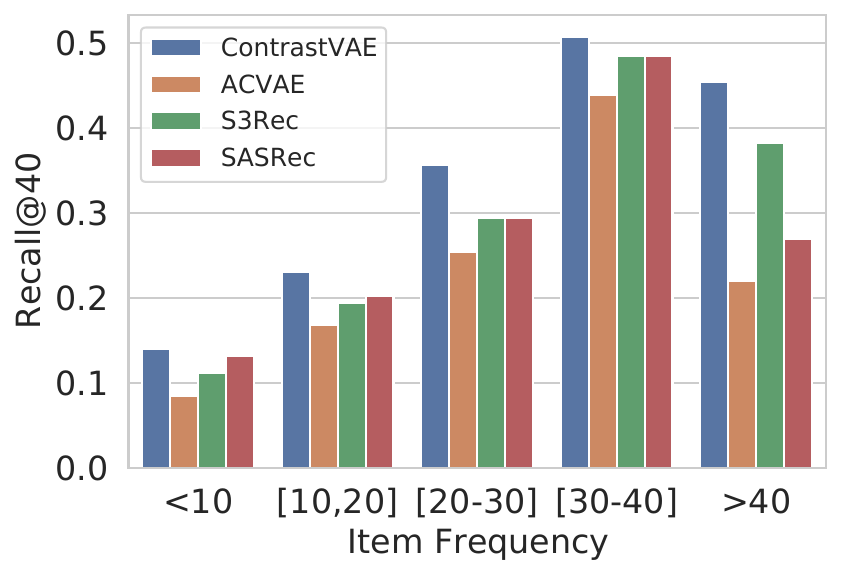} 
\caption{Item frequencies}
\label{fig:sub_frequency}
\end{subfigure}
\begin{subfigure}{0.48\columnwidth}
\includegraphics[width=\linewidth]{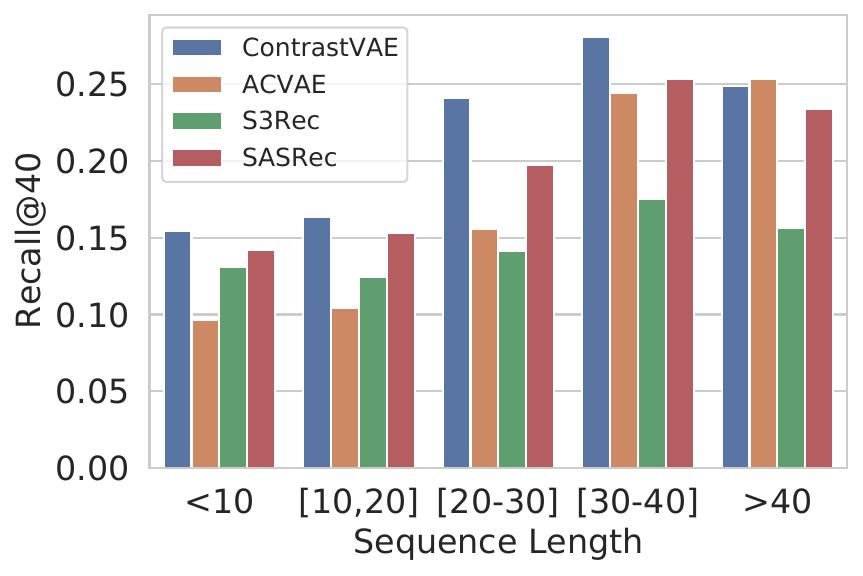}
\caption{Sequence lengths}
\label{fig:sub_length}
\end{subfigure}
\caption{Subgroup sequences analysis on Toy dataset}
\label{fig:image5}
\end{figure}

We first investigate \modelname's performance on long-tail items, which have much fewer interactions in the training set and thus are much harder to predict. To reach this target, we follow the grouping strategies in Table~\ref{tab:data_freq_per_item} and categorize the user sequences into $5$ groups according to the frequencies of their last clicked items. We report Recall@40 of \modelname and representative baseline methods on the Toy dataset in Fig.~\ref{fig:sub_frequency}. We can observe that \modelname achieves the highest Recall@40 sores on all groups of sequences. Specifically, on long-tail items (i.e., $[\le 10]$ and $[10, 20]$), our method outperforms other baseline models by a large margin. The improvement over such user sequences greatly contributes to the overall performance of our method, as most items to be predicted are long-tail items.  

We further study how the sequence length affects the model's performance. Similar to the item frequency, we split user sequences into $5$ groups according to their lengths, and we report the performance of \modelname and other models on the Toy dataset in Fig.~\ref{fig:sub_length}. It demonstrates that \modelname consistently exhibits good performance on sequences with various lengths, which highlights the effectiveness of our model. Similar to the results on long-tail items, \modelname greatly improves the performance of short sequences (i.e., user sequences which have less than 20 interactions). However, on longer sequences (e.g., $[\ge 40$] ), \modelname does not show superior performance compared with other models, and we guess that for long sequences, the users' preferences tend to become certain and easy to predict, in which case the uncertainty and randomness introduced by our method would not help the prediction results.

\subsubsection{\modelname alleviates posterior collapse and point-estimation in latent space}
We then study why \modelname can improve VAEs for sequential representation through analyzing the posterior distributions learned from \modelname and the vanilla VAE model AVAE. 
We adopt two metrics to evaluate the quality of learned latent variables $\bm{z}$: 1) the average KL divergence between the posterior distribution $p(\bm{z}|\bm{x})$ of sequences and the standard Gaussian distribution $\mathcal{N}(\bm{0}, \mathbf{I})$, which reflects the extent of posterior collapse problem (posterior collapse induces small KL-divergence); 2) the average variance of latent variables, which reflect the extent of variance vanishing. 

\begin{figure}[h]
\begin{subfigure}{0.23\textwidth}
\includegraphics[width=\linewidth,height=3cm]{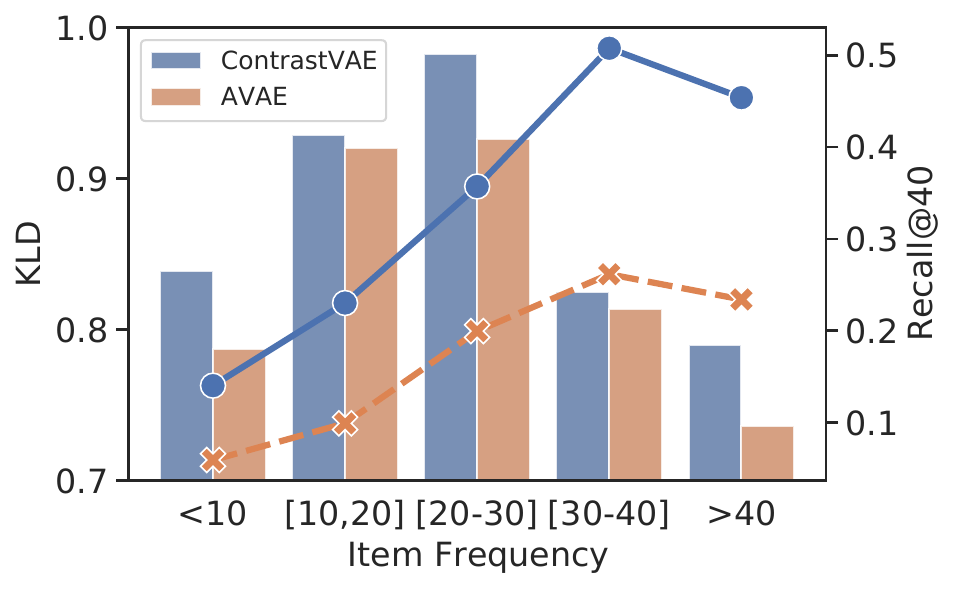}
\caption{Recall@40 (line graph) and KL-divergence (bar graph)}
\label{fig:subim4}
\end{subfigure}
\begin{subfigure}{0.23\textwidth}
\includegraphics[width=\linewidth,height=3cm]{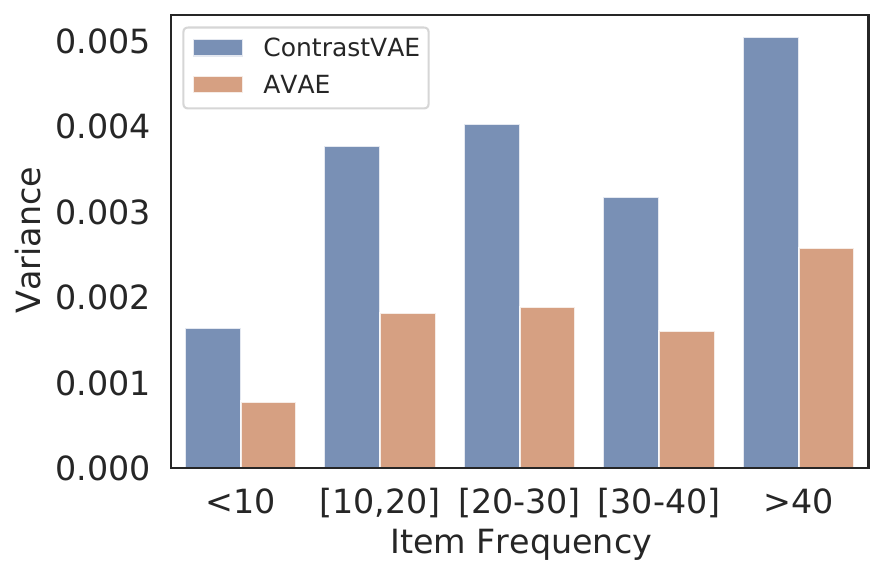}
\caption{Variance of latent variable estimation}
\label{fig:subim3}
\end{subfigure}
\caption{Latent space analysis on the Toy dataset.}
\label{fig:cl_latent}
\end{figure}

We report the Recall@40 scores and these two metrics of sequences targeted items with different frequencies in Fig.~\ref{fig:cl_latent}. For AVAE, the KL-divergences between latent variable estimation and standard Gaussian distribution are very low, especially for infrequent items, indicating that they suffer from posterior collapse problems. Then, in this case, the representations of sequences that target different items would be hard to  discriminate in latent space, leading to poor performance of AVAE model on infrequent items. By contrast, \modelname alleviates the posterior degeneration through CL, which implicitly encourages different sequences to have diverse latent representations. An interesting observation is that the KL-divergence of both \modelname and AVAE decreases when the item's frequency gets extremely large (e.g., $>30$), and we guess it is because popular items are much easier to predict, and thus, the model does not require a fine-grained posterior estimation to model it. From Fig.~\ref{fig:subim3} we notice that AVAE model has vanishing variance over infrequent items, which indicates that AVAE collapses to point estimation for such infrequent items. On the contrary, \modelname managed to increase the average variance over such sequences, thus increasing the robustness of noise.

\subsubsection{Robustness analysis.}
\begin{figure}[h]
\begin{subfigure}{0.48\columnwidth}
\includegraphics[width=\linewidth]{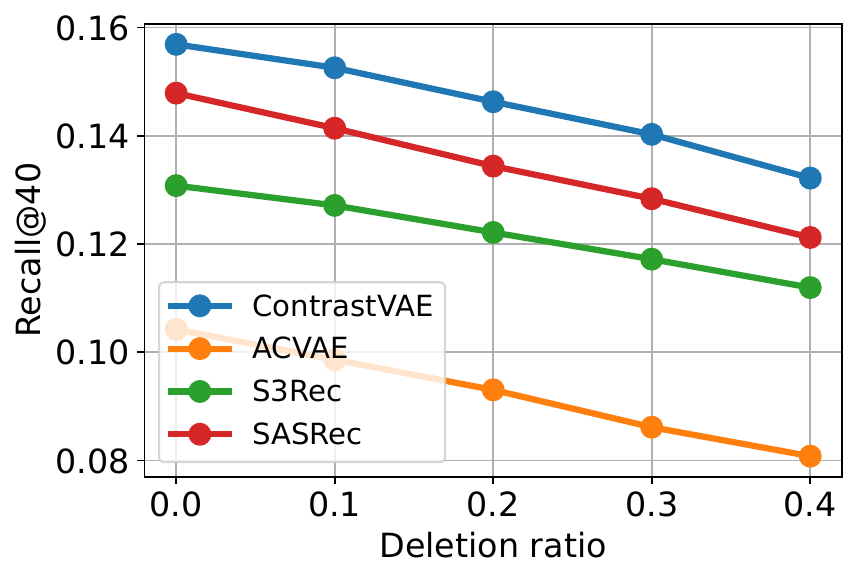} 
\caption{Random deletion}
\label{fig:robustness_deletion}
\end{subfigure}
\begin{subfigure}{0.48\columnwidth}
\includegraphics[width=\linewidth]{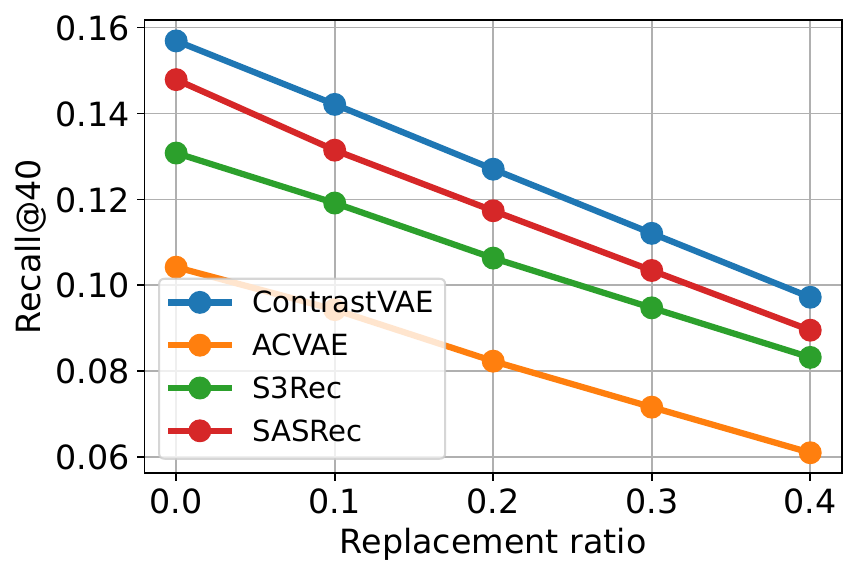}
\caption{Random replacement}
\label{fig:robustness_replacement}
\end{subfigure}
\caption{Robustness analysis on the Toy dataset.}
\label{fig:robustness}
% \label{fig:}
\end{figure}

We further study the impact of corrupted input sequences for \modelname to analyze its robustness w.r.t. to noisy data. We consider two corrupting strategies: 1) randomly deleting a proportion of items in each sequence (random deletion); 2) randomly replacing proportion items with other items in each sequence (random replacement). Consistent with Sec.~\ref{sec:study1}, we compare our method with typical baseline models ACVAE, S3Rec, and SASRec. As shown in Fig.~\ref{fig:robustness}, the performance of all models exhibits a drop as we increase the corruption ratio. However, \modelname always outperforms other baseline models by a large margin whatever the corruption method and the corruption ratio, which indicates that \modelname can still exhibit good performance for noisy input data.
\vspace{-10pt}
\subsubsection{Hyper-parameter sensitivity analysis}
\begin{figure}[h] 
\begin{subfigure}{0.48\columnwidth}
\includegraphics[width=\linewidth]{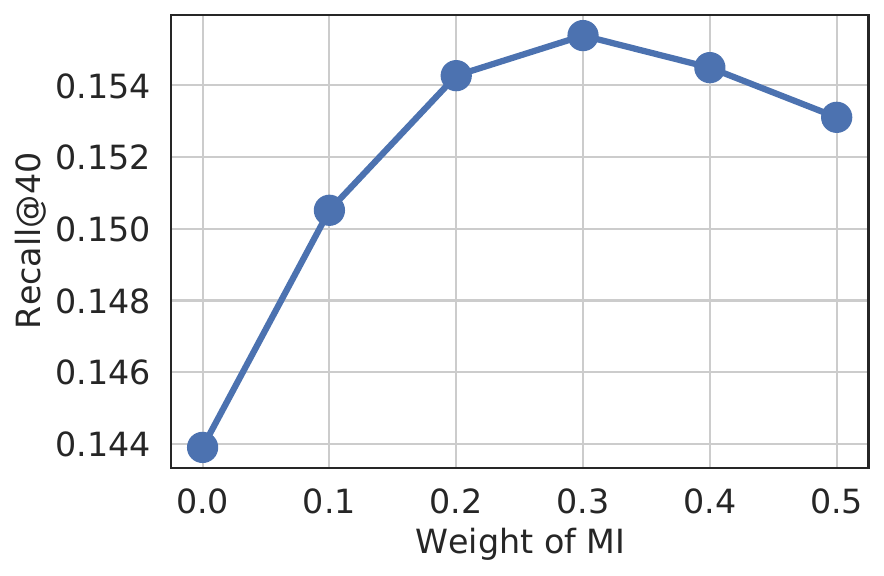} 
\caption{Beauty}
\label{fig:sense-beauty}
\end{subfigure}
\begin{subfigure}{0.48\columnwidth}
\includegraphics[width=\linewidth]{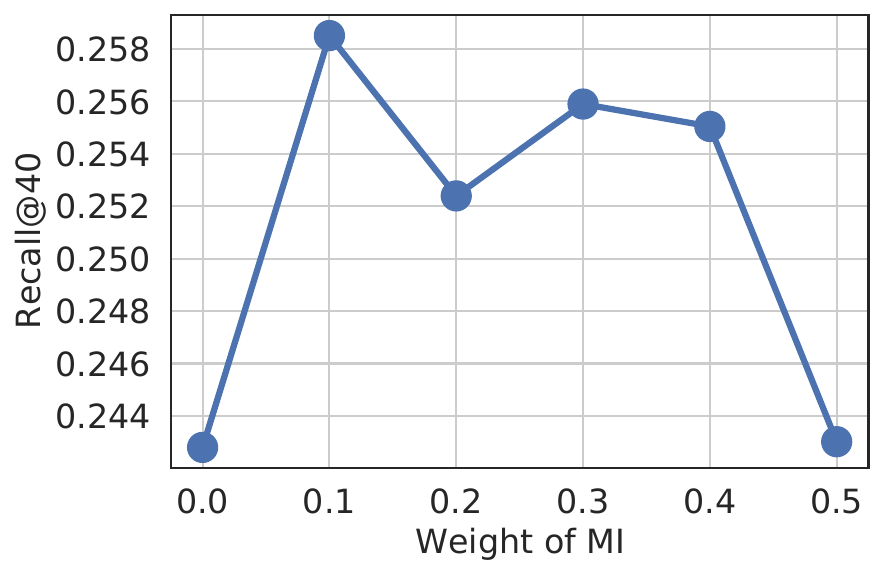}
\caption{Office}
\label{fig:sense-office}
\end{subfigure}
\caption{Sensitivity analysis of the weight of contrastive loss.}
\label{fig:sensitivity}
\end{figure}

We finally study the performance variation of our model w.r.t. the intensity of the CL by tuning the weight of MI loss $\lambda$, and we present the Recall@40 scores on Beauty and Office dataset in Fig.~\ref{fig:sensitivity}. We find that compared with not using mutual information (i.e., the weight $\lambda$ is set as $0$), a proper weight can lead to great improvements (up to $7.6\%$). Also, the weight cannot be too large otherwise, it would constrain the model's learning from the next-item prediction tasks.
\section{Conclusion}
In this paper, we have proposed \modelname, a novel method for SR. We start by extending single-view evidence lower-bound to two-view cases and derive ContrastELBO. Then to optimize ContrastELBO for SR tasks, we propose \modelname, which takes two views of sequential data as input and optimizes an additional mutual information maximization term besides conventional reconstruction loss and kl-divergence loss of two views. We further propose model augmentation and variational augmentation for generating another view for an input sequence to solve the inconsistency problem led by conventional data augmentation methods. Experiment results and analysis show that our architecture combined with augmentations outperforms other competitive baselines.

\section{ACKNOWLEDGEMENTS}
This work is supported in part by NSF under grants III-1763325, III-1909323,  III-2106758, and SaTC-1930941. 

\bibliographystyle{ACM-Reference-Format}
\bibliography{references}
\end{document}